# Title: Deformation of an inner valence molecular orbital in ethanol by an intense laser field

**Authors:** Hiroshi Akagi[1]*, Tomohito Otobe[1], Ryuji Itakura[1]

**Affiliations:**

[1] Kansai Photon Science Institute, National Institutes for Quantum and Radiological Science and Technology (QST), 8-1-7 Umemidai, Kizugawa, Kyoto 619-0215, Japan

*Corresponding author. Email: akagi.hiroshi@qst.go.jp

**Abstract:** Valence molecular orbitals play a crucial role in chemical reactions. Here we reveal that an intense laser field deforms an inner valence orbital (10a') in the ethanol molecule. We measure the recoil-frame photoelectron angular distribution (RFPAD), which corresponds to the orientation dependence of the ionization probability of the orbital, using photoelectron-photoion coincidence momentum imaging with a circularly polarized laser pulse. *Ab initio* simulations show that the orbital deformation depends strongly on the laser field direction, and that the measured RFPAD cannot be reproduced without taking the orbital deformation into account. Our findings suggest that the laser-induced orbital deformation occurs prior to electron emission on a sub-optical-cycle time scale.

**One Sentence Summary:** Orbital deformation on a sub-optical-cycle time scale is revealed by orientation dependence of tunnel ionization.

**Main Text:** Molecular orbital (MO) shape plays a crucial role in chemical reaction dynamics and is described by theories such as the frontier orbital (*1*) and Woodward–Hoffmann rules (*2*). Similarly, in intense-laser-induced molecular dynamics, MO shape is known to be important particularly in tunnel ionization (*3, 4*). Tunnel ionization is a crucial initial step of successive dynamical processes leading to high-harmonic generation, dissociative ionization, and multiple ionization. In the last two decades, orientation dependence of tunnel ionization probabilities has been studied mainly for small molecules composed of two or three atoms (*5-7*). These studies have shown that the shape of an MO from which an electron tunnels strongly affects the orientation dependence of the ionization probability. Hence, the ionization probability as a function of molecular orientation is measured to image the MO shape. This molecular scanning tunnel microscopy (STM) (*8*) has revealed that laser-induced tunneling occurs not only from the highest-occupied molecular orbital (HOMO), but also from inner valence orbitals (*9*). Inner valence orbitals play a crucial role in intense-field chemistry.

In addition to contribution of multiple MOs (*9-11*), laser-induced deformation of HOMOs has been studied in tunnel ionization. Simulations including HOMO deformation agree well with the measured angular dependence of ionization probability for $CO_2$ (*12*) and with high-harmonic spectra from $N_2$ (*13*). Experimentally, high-harmonic spectroscopy with the aid of *ab initio* simulation has revealed that degenerate HOMOs in spatially oriented $CH_3F$ are deformed by intense laser fields, which remove the degeneracy (*14*). The contribution of unoccupied MOs such as Rydberg orbitals was also suggested for dissociative ionization of hydrocarbon molecules (*15*). There is also evidence that intense laser fields modify the electronic structure of ionized molecules (*16, 17*). In this study, we demonstrate laser-induced MO deformation of an inner valence orbital in a neutral molecule. Photoionization from an inner valence orbital is essential to creating electronically excited ions, which is necessary for dissociation (*18-21*). For efficient control of laser-driven dissociative ionization, it is important to deepen our understanding of electronic dynamics of inner valence orbitals.

Ethanol is an ideal molecule for exhibiting MO deformation in a strong laser field. Ethanol has rather low symmetry (the point group Cs) with its symmetry plane including an O atom and two C atoms as shown in Fig. 1A. Hence, all MOs of ethanol are classified in two irreducible representations: a′ and a″. The energy levels of the four inner valence MOs, 10a' (HOMO-1), 2a" (HOMO-2), 9a' (HOMO-3), and 8a' (HOMO-4), lie within 3 eV of each other in the field-free situation (*22*), and three of the four MOs have a' symmetry, as illustrated in the inset of Fig. 1A. In our experiment, we used a circularly polarized laser field with an intensity of $I_0 = 8\times10^{13}$ W/cm$^2$. The electric field of $E = 1.7\times10^{10}$ V/m creates a slope with an energy difference of about 6 eV for a distance of ~ 3.5 Å, which corresponds to the size of ethanol (Fig. 1A). This difference is larger than the energy range of the three a' MOs in the field-free situation. Therefore, the laser field should induce strong mixing of these MOs. Additionally, a previous photoelectron-photoion coincidence (PEPICO) measurement with a He lamp (He I at 21.2 eV) showed that electron emission from the inner valence MOs results in the formation of different fragment ions (*23*). This means that identifying fragment ion species allows us to identify the MO that has an electron hole just before the ethanol cation dissociates (*18, 19, 21*).

First, we show theoretically that 10a' (HOMO-1) in ethanol is deformed by a circularly polarized intense laser field ($I_0 = 8\times10^{13}$ W/cm$^2$, $E = 1.7\times10^{10}$ V/m) during one cycle of the electric (***E***) field (Fig. 1A). The field-free MO 10a' ($\mathbf{\Psi}_{10a'}$), shown at the center of Fig. 1A, is composed mainly of the lone pair on the O atom and the C-C $\sigma$ bond. We calculate the 10a' orbitals in the ***E***-field with density functional theory (DFT) [supplementary materials (SM) Section 2]. The deformed orbital shapes $\mathbf{Y}_{10a'}(\Phi)$ are drawn around field-free $\mathbf{\Psi}_{10a'}$ in Fig. 1A, where the ***E***-field is parallel to the Cs symmetry plane with the angle $\Phi$ between the ***E***-field direction and the C-C axis. In the direction $\Phi = -22.5°$ (the CH$_3$ side), $\mathbf{Y}_{10a'}(\Phi = -22.5°)$ is similar to field-free 9a' ($\Psi_{9a'}$) shown in Fig. 1B. In the other directions, the deformed MO $\Upsilon_{10a'}(\Phi)$ exhibits only slight differences from field-free $\Psi_{10a'}$. Nevertheless, $\Upsilon_{10a'}(\Phi)$ expands in the direction opposite to the *E*-field. These deformations of 10a' occur on a sub-optical-cycle time scale of the circularly polarized laser pulse, as illustrated in Fig. 1A. The field-deformed MO $\Upsilon_{10a'}(\Phi)$ can be described as a linear combination of the neighboring a' MOs, $\Psi_{10a'}$ (HOMO-1), $\Psi_{9a'}$ (HOMO-3), and $\Psi_{8a'}$ (HOMO-4), shown in Fig. 1B.

The deformation from $\Psi_{10a'}$ to $\Upsilon_{10a'}(\Phi)$ in the *E*-field affects angular dependence of the ionization probability. In our experiment, we applied molecular STM to ethanol in a circularly polarized laser field ($\lambda \sim 795$ nm, $\Delta\tau \sim 60$ fs, $I_0 \sim 8\times10^{13}$ W/cm$^2$) (*9, 24*). We used a partially deuterated ethanol sample, CH$_3$CD$_2$OH, to avoid ambiguity in the mass assignment caused by producing different fragment ions with the same mass. An unaligned ethanol molecule was singly ionized in the circularly polarized intense laser field, and three-dimensional momentum vectors of the electron ($\vec{p}_{\text{ele}}$) and ion ($\vec{p}_{\text{ion}}$) produced were measured in coincidence with two position-sensitive detectors (Fig. 2A). Here, we focused on CD$_2$OH$^+$ (m/z = 33) production because the previous PEPICO measurement with a He lamp found that CD$_2$OH$^+$ is produced by electron emission from the 10a' MO (HOMO-1) (*23*). In the circularly polarized intense laser field, the freed electron drifts perpendicularly to the *E*-field direction at the moment of tunneling (*25*). The recoil direction of the CD$_2$OH$^+$ ion reflects the orientation of the parent molecule just before dissociation. Thus, the recoil-frame photoelectron angular distribution (RFPAD) for the CD$_2$OH$^+$ production channel in the circularly polarized laser field is derived with respect to the CD$_2$OH$^+$ recoil vectors (SM Section 1). The RFPAD shows preferential electron tunneling from the CH$_3$ side of CH$_3$CD$_2$OH (Fig. 2B).

To understand the experimental RFPAD, we simulated the angular dependence of the tunnel ionization probability using DFT (*4*) (SM Section 2). Figure 3A is the simulated ionization probability $W_{10a'}(\Phi, \Theta)$ of $\Upsilon_{10a'}(\Phi, \Theta)$ as a function of the *E*-field direction $(\Phi, \Theta)$ defined in the inset of Fig. 3A. The two-dimensional map of $W_{10a'}(\Phi, \Theta)$ shows the maximum at $(\Phi, \Theta) = (157.5°, 90°)$. In other words, an electron in 10a' tunnels preferentially from the $CH_3$ moiety as illustrated in Fig. 3D; thus, it is somewhat similar to the RFPAD measured for the $CD_2OH^+$ channel (Fig. 2B).

To evaluate the effect of MO deformation on tunnel ionization, we simulated the ionization probability $W^0_{10a'}(\Phi, \Theta)$ of the field-free MO $\Psi_{10a'}$, as shown in Fig. 3B (SM Section 3). Comparing $W^0_{10a'}(\Phi, \Theta)$ with $W_{10a'}(\Phi, \Theta)$ clarifies the effect of MO deformation. The ionization probability $W^0_{10a'}(\Phi, \Theta)$ of field-free $\Psi_{10a'}$ has its maximum at $(\Phi, \Theta) = (-112.5°, 90°)$, where the electron tunnels from the lone pair on the O atom. The probability $W_{10a'}(\Phi, \Theta)$ of $\Upsilon_{10a'}(\Phi, \Theta)$ in the *E*-field does not have any peaks around the direction $(\Phi, \Theta) = (-112.5°, 90°)$ (green dashed square in Fig. 3A), but has its maximum at $(\Phi, \Theta) = (157.5°, 90°)$ (pink dashed square), corresponding to electron tunneling from the $CH_3$ moiety. The change in direction of maximum ionization probability can be explained by the significant contribution of $\Psi_{8a'}$ to $\Upsilon_{10a'}(\Phi, \Theta)$, as presented in Figs. 3C and D. The MO $\Upsilon_{10a'}(\Phi, \Theta)$ in the *E*-fields pointing to $(\Phi, \Theta) = (-112.5°, 90°)$ and $(157.5°, 90°)$ can be expressed as a linear combination of $\Psi_{10a'}$ and $\Psi_{8a'}$, which enlarges the lobe around the $CH_3$ moiety owing to the constructive overlap and shrinks the lone pair on the O atom because of the destructive overlap. The deformation of 10a' enhances electron tunneling from the $CH_3$ moiety and suppresses that from the lone pair on the O atom, as shown in Figs. 3C and D.

The experimental RFPAD for the $CD_2OH^+$ channel (Fig. 2B) gives evidence of MO deformation. To compare the experimental and simulated results, we derive the theoretical RFPADs from the two-dimensional maps of $W_{10a'}(\Phi, \Theta)$ and $W_{10a'}^0(\Phi, \Theta)$ shown in Figs. 3A and B. Here, the electron is assumed to tunnel in the opposite direction of the **E**-field with an angular uncertainty of ± 16°, which is estimated from the out-of-plane photoelectron angular distribution with respect to the polarization plane (SM Section 4). For the fragment ion emission, we apply the axial recoil approximation. The RFPAD $\Omega_{10a'}(\varphi_{RFPAD})$ derived from $Y_{10a'}(\Phi, \Theta)$ in the **E**-field has a peak at $\varphi_{RFPAD} \sim 180°$ (green line in Fig. 4A), showing the preferential electron tunneling from the $CH_3$ moiety. This agrees with the experimental RFPAD (pink line in Fig. 4A). Nevertheless, the theoretical RFPAD has a minimum at $\varphi_{RFPAD} \sim 90°$, which we do not observe in the experimental RFPAD. We attribute this discrepancy to the electron tunneling from 3a" (HOMO), followed by the subsequent electronic excitation to the first electronically excited $1\ ^2A'$ state [an electron hole in 10a' (HOMO-1)] of the ethanol cation. This stepwise process also produces the $CD_2OH^+$ ion as illustrated in Fig. 4C (*18, 19, 21*).

To describe electronic excitation in intense laser fields, perturbative photo-absorption is not appropriate. As illustrated in Fig. 1A, the molecular orbital varies adiabatically as the **E**-field rotates. Electronic excitation in intense laser fields can be described as non-adiabatic transition between the laser-driven adiabatic electronic states (*26, 27*). Thus, non-adiabatic excitation following electron tunneling plays a key role. Because of the two different pathways producing $CD_2OH^+$, the experimental RFPAD should be expressed as a linear combination of two pathways from 10a' and 3a",

$$\Omega_{CD_2OH^+}^{exp}(\varphi_{RFPAD}) \propto \Omega_{10a'}(\varphi_{RFPAD}) + f_{3a"}^{exc}\Omega_{3a"}(\varphi_{RFPAD}), \tag{1}$$

where $f_{3a''}^{exc}$ is the excitation fraction to the first excited 1 $^2$A' state. The fraction $f_{3a''}^{exc}$ is calculated by using the measured ion yields of CD$_2$OH$^+$, CH$_3$CD$_2$OH$^+$ (m/z = 48), and CH$_3$CDOH$^+$ (m/z = 46) (Fig. 4C). The CD$_2$OH$^+$ yield $I(CD_2OH^+)$ is expressed as $I(CD_2OH^+) \propto P_{10a'} + f_{3a''}^{exc} P_{3a''}$, where $P_{10a'}$ and $P_{3a''}$ are the integrated tunneling probabilities from 10a' and 3a'' (SM Section 4). The sum of the CH$_3$CD$_2$OH$^+$ and CH$_3$CDOH$^+$ yields, both of which are correlated to the electronic ground 1 $^2$A'' state of CH$_3$CD$_2$OH$^+$ (23), is similarly expressed as $I(CH_3CD_2OH^+) + I(CH_3CDOH^+) \propto (1 - f_{3a''}^{exc}) P_{3a''}$. Consequently, we can express the fraction $f_{3a''}^{exc}$ as

$$f_{3a''}^{exc} = (R_{ion} P_{3a''} - P_{10a'})/(1 + R_{ion}) P_{3a''}, \qquad (2)$$

where $R_{ion} = I(CD_2OH^+)/[I(CH_3CD_2OH^+) + I(CH_3CDOH^+)]$. Inserting the measured ion yield ratio $R_{ion}$ = 1.3, we obtain $f_{3a''}^{exc}$ = 0.48. Using the obtained $f_{3a''}^{exc}$, the linear combination of eq. (1) reasonably agrees with the experiment as shown in Fig. 4D. Whereas, the RFPADs from the field-free MOs ($\Psi_{10a'}$ and $\Psi_{3a''}$) show a peak at around $\varphi_{RFPAD}$ = 80° (green and blue lines in Fig. 4B), and the measured peak at $\varphi_{RFPAD}$ ~ 180° (pink line) cannot be reproduced by linear combination of the two RFPADs from $\Psi_{10a'}$ and $\Psi_{3a''}$. This suggests that MO deformation is needed to explain the experimental RFPAD for the CD$_2$OH$^+$ channel. As shown in Fig. 1B, the contribution of the field-free 8a' orbital to the deformed HOMO-1 orbital is essential for the peak at $\varphi_{RFPAD}$ ~ 180°. The remaining discrepancy between the theory and experiment in Fig. 4D is attributed to the deviation from the axial-recoil approximation in the CD$_2$OH$^+$ production.

We have shown that an inner valence MO (10a') in ethanol is deformed by a circularly polarized intense laser field on the sub-optical-cycle time scale, resulting in an observed orientation dependence of the ionization probability. The MO deformation is not unique to ethanol but should occur in general, especially for polyatomic molecules with a high density of inner valence levels. The orientation dependence of the MO deformation presented in this study will open the door to direct control of electronic dynamics leading to selective bond breaking. Combining molecular orientation and a sub-cycle probe would serve this end.

**Acknowledgments:** We thank K. Hosaka, K. Yamanouchi, and A. Yokoyama for support in developing the PEPICO apparatus; Y. Hagihara for support in data analysis; and H. Kono and M. Tsubouchi for valuable discussions. **Funding:** We acknowledge valuable financial support from JSPS KAKENHI Grant Numbers JP22685004, JP23350013, JP26288013, and JP17H03525.
**Author contributions:** H.A. and R.I. conceived and designed this study. R.I. developed the laser system and PEPICO apparatus. H.A. conducted the experiments and analyzed the measured data. T.O. performed the *ab initio* calculations and provided the theoretical interpretation. R.I. supervised the project. H.A. drafted the original manuscript, and all authors edited and reviewed the manuscript. **Competing interests:** The authors declare no competing financial interests.
**Data and materials availability:** All data are available in the main text or the supplementary materials.

**Supplementary Materials:**

Materials and Methods

Figures S1-S5

Tables S1

References (*28-35*)

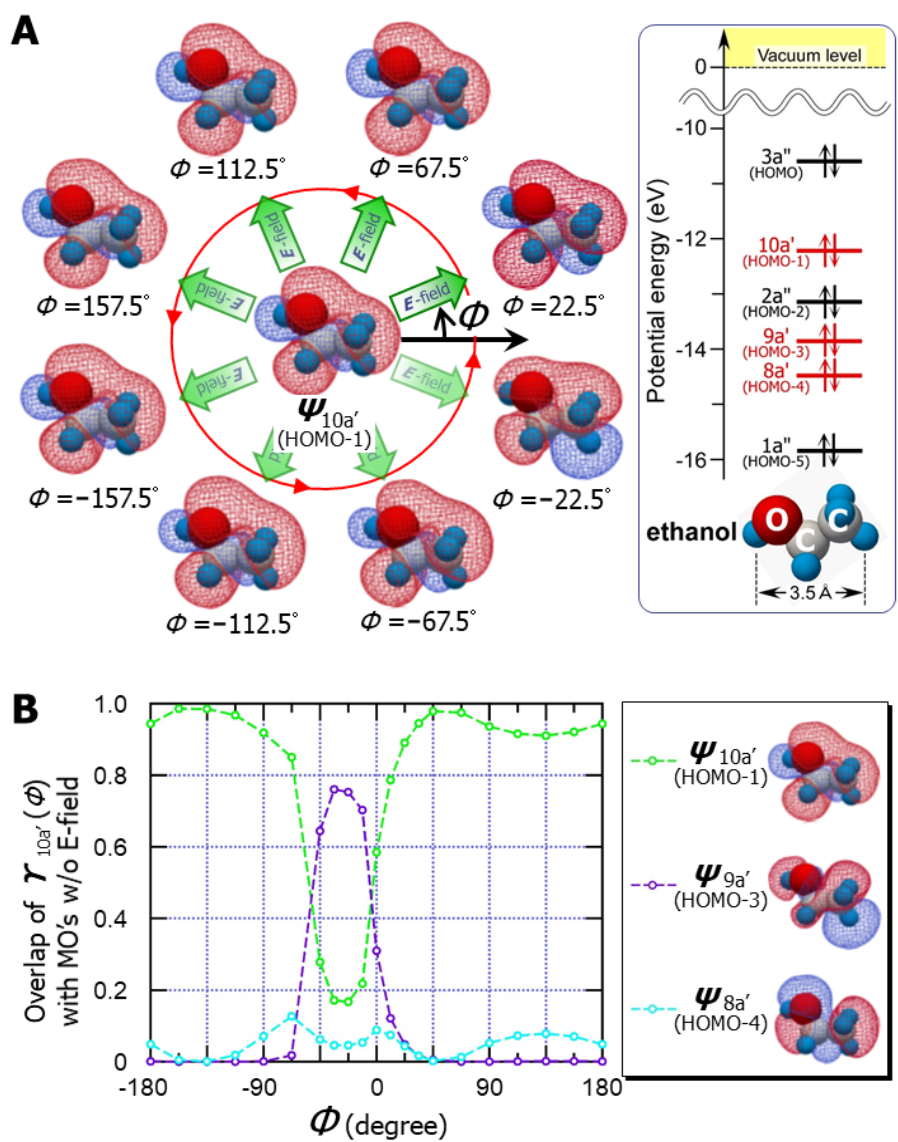

**Fig. 1. MO deformation induced by a laser electric field.**

**(A)** DFT-calculated 10a' (HOMO-1) structures of ethanol in an electric field with strength of $1.7\times10^{10}$ V/m (corresponding to a circularly polarized laser field at an intensity of $8\times10^{13}$ W/cm$^2$) as a function of field direction. The electric field direction is set parallel to the Cs symmetry plane and defined by the angle $\Phi$ from the C-C axis of ethanol. Isosurface plots of the MOs $\Upsilon_{10a'}(\Phi)$ in the electric field with cutoff values of 0.025 (blue) and – 0.025 (red) are drawn around that of the field-free MO ($\Psi_{10a'}$). **Inset:** Energy level diagram of the field-free MOs in ethanol. The red and black levels have a' and a" symmetries, respectively. **(B)** Overlap populations $|\langle\Psi_i|\Upsilon_{10a'}(\Phi)\rangle|^2$ of $\Upsilon_{10a'}(\Phi)$ with the field-free MOs $\Psi_{8a'}$, $\Psi_{9a'}$, and $\Psi_{10a'}$, as functions of angle $\Phi$.

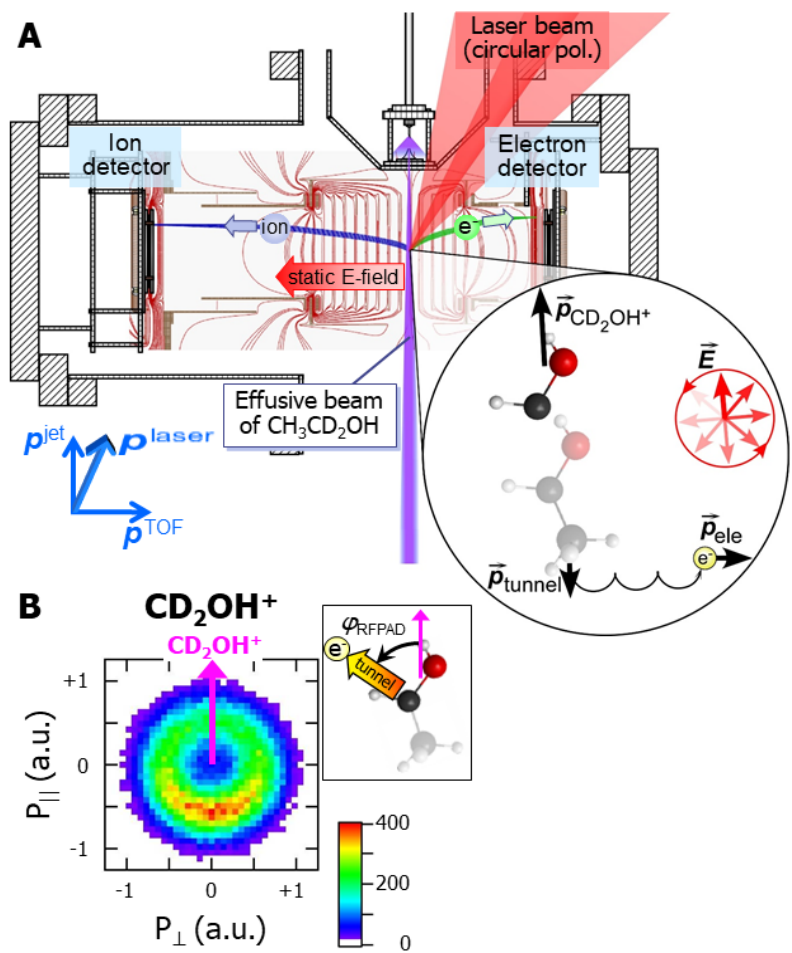

**Fig. 2. Recoil-frame photoelectron momentum measurements.**

**(A)** Sketch of relations between the electron tunneling direction ($\vec{p}_{tunnel}$), final photoelectron momentum ($\vec{p}_{ele}$), recoil momentum of the $CD_2OH^+$ ion ($\vec{p}_{CD_2OH^+}$), and $E$-field direction ($\vec{E}$) of the circularly polarized laser field in the experimental setup. **(B)** Recoil-frame photoelectron momentum distribution for the $CD_2OH^+$ channel. The arrow in the image indicates the recoil direction of the fragment ion.

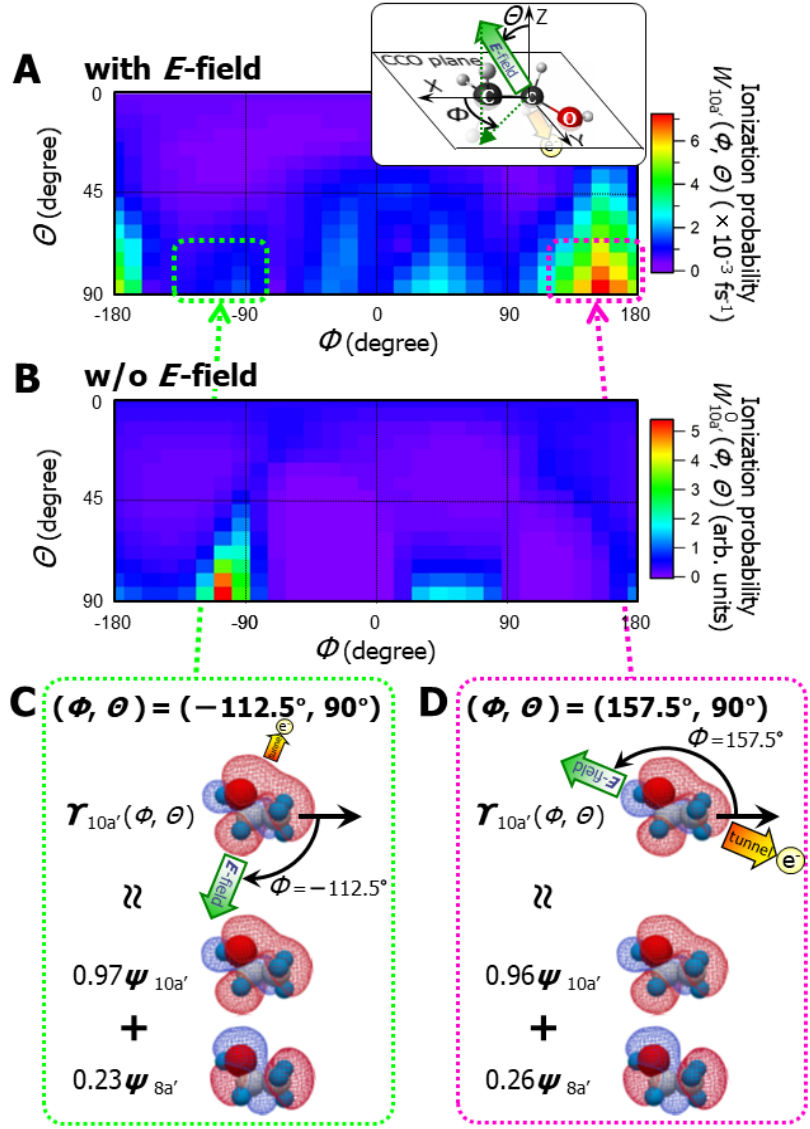

**Fig. 3. MO deformation effect on the angular-dependent ionization probability.**

**(A)** DFT-calculated angular-dependent ionization probability of the 10a' MO [$\Upsilon_{10a'}(\Phi, \Theta)$] in the electric field with $E = 1.7 \times 10^{10}$ V/m (SM Section 2). **Inset:** Defined electric field direction represented with Euler angles $\Phi, \Theta$. **(B)** Same as (A) but simulated for the field-free 10a' MO ($\Psi_{10a'}$) (SM Section 3). **(C and D)** Field-deformed MOs $\Upsilon_{10a'}(\Phi, \Theta)$ represented by linear combinations of the field-free MOs $\Psi_{10a'}$ and $\Psi_{8a'}$ at $(\Phi, \Theta) = (-112.5°, 90°)$ and $(157.5°, 90°)$, respectively.

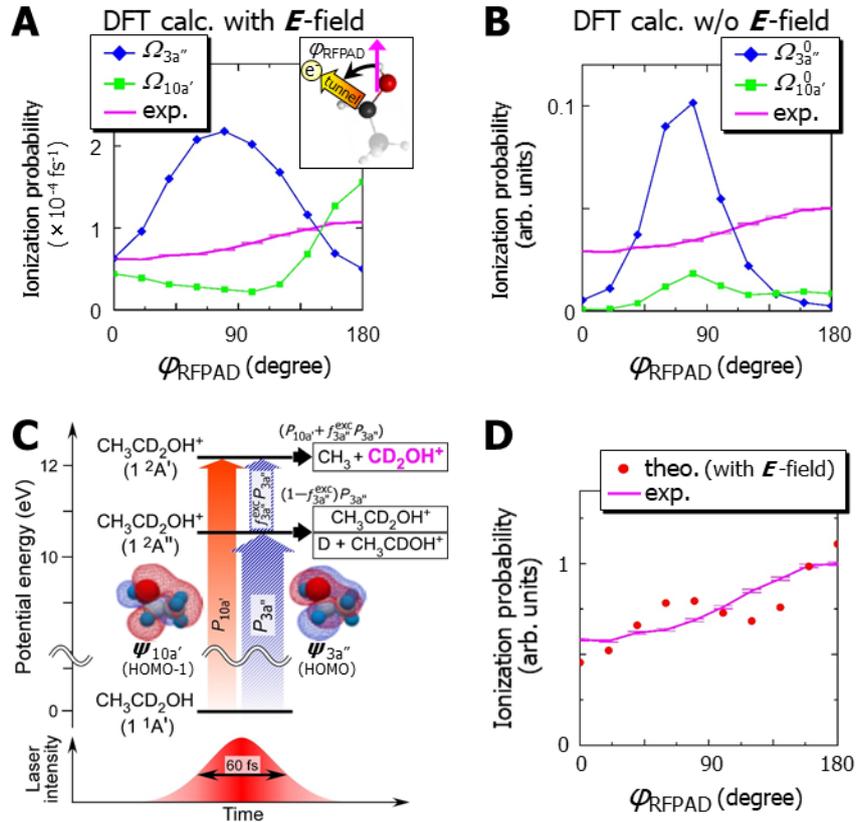

**Fig. 4. Comparison of theoretical and experimental RFPADs.**

**(A)** Theoretical RFPADs $\Omega_{3a''}(\varphi_{\text{RFPAD}})$ and $\Omega_{10a'}(\varphi_{\text{RFPAD}})$ from field-deformed $\Upsilon_{3a''}(\Phi,\Theta)$ and $\Upsilon_{10a'}(\Phi,\Theta)$, respectively, in the electric field with $E = 1.7\times 10^{10}$ V/m (see SM Section 4 for details). The experimental RFPAD is also shown with its vertical error bars. **(B)** Same as (A) but simulated for the field-free MOs. **(C)** Schematic for the tunnel ionization and subsequent processes of $CH_3CD_2OH$ in the circularly polarized laser field. **(D)** Comparison of the experimental RFPAD with the linear combination of the theoretical RFPADs from the field-deformed MOs [eq. (1) with $f_{3a''}^{\text{exc}} = 0.48$].

17